\documentclass[tightenlines,superscriptaddress,eqsecnum,floats,nofootinbib,showpacs]
{revtex4}
\usepackage{amssymb}
\usepackage{stmaryrd}
\usepackage{amsmath}
\usepackage{amsfonts}
\usepackage{mathrsfs}
\usepackage{amsmath,amssymb,amsfonts}
\usepackage{graphicx}
\usepackage[T1]{fontenc}

\def\be{\begin{equation}}
\def\ee{\end{equation}}
\def\ba{\begin{eqnarray}}
\def\ea{\end{eqnarray}}
\def\nn{\nonumber}

\newcommand{\mubar}{{\bar \mu}} 

\newcommand{\sgn}{\mathrm{sgn}} 
\newcommand{\dd}{{\rm d}}
\newcommand{\tbc}{T_{\rm bnc}^{(c)}}
\newcommand{\tbb}{T_{\rm bnc}^{(b)}}
\newcommand{\tbbpm}{T_{\rm bnc}^{(b\pm)}}
\newcommand{\tbbp}{T_{\rm bnc}^{(b-)}}
\newcommand{\tbbm}{T_{\rm bnc}^{(b+)}}

\newcommand{\arpx}{{\rm Ar}(\square(\phi,x))}
\newcommand{\artp}{{\rm Ar}(\square(\theta,\phi))}

\begin{document}


\title{ Quantum geometry and effective dynamics of Janis-Newman-Winicour singularities }

\author{Cong Zhang\footnote{ zhang.cong@mail.bnu.edu.cn}}
\affiliation{ Faculty of Physics, University of Warsaw, Pasteura 5, 02-093 Warsaw, Poland}

\author{Xiangdong Zhang\footnote{Corresponding author: scxdzhang@scut.edu.cn}}
\affiliation{Department of Physics, South China University of
Technology, Guangzhou 510641, China}

\begin{abstract}
Inspired by the recent proposal for the quantum effective dynamics of the Schwarzschild spacetime given in \cite{AOS1}, we investigate the effective dynamics of the loop quantized Janis-Newman-Winicour (JNW) spacetime which is an extension of the Schwarzschild spacetime with an extra minimally coupled massless scalar field. Two parameters are introduced in order to regularize the Hamiltonian constraint in the quantum effective dynamics. These two parameters are assumed to be Dirac observables when the effective dynamics is solved. By carefully choosing appropriate conditions for these two parameters, we completely determine them, and the resulted new effective description of the JNW spacetime leads to a well behaved quantum dynamics which on one hand resolves the classical singularities, and on the other hand, agrees with the classical dynamics in the low curvature region.
\end{abstract}


\maketitle

\section{Introduction}
It is well known that Einstein's theory of general relativity (GR) fails to give any concrete predictions around singularities due to the non-negligible quantum effects in the Planck regime.
Various attempts has been made to construct a consistent theory describing such a regime in the past decades. Among these theories, loop quantum gravity (LQG) presents a picture of granular and discrete space-time at Planck scale\cite{Ro04,Th07,As04,Ma07}. In this theory, it has been shown that the operators representing geometric observables (e.g., 2-surface area, 3-region volume, length of a curvature and integral of certain metric components) have a discrete spectra \cite{Ro04,Th07}. Namely, gravity in LQG is quantized. Because of the quantum features of the underlying spacetime geometry, LQG propose that singularities may not exist. This concept was first implemented precisely in the theory of loop quantum cosmology (LQC) which is constructed by applying the method of loop quantization to the homogeneous and isotropic cosmological model \cite{LQC5,Boj,Ash-view,AS11,BCM}. According to LQC, the classical big bang singularity is finally resolved by the quantum bounce scenario \cite{APS3,ACS}.

Beyond the big-bang singularity, another well-known type of singularities is inside the black holes. The simplest example is the Schwarzschild interior space-time. Because this region is isometric to the Kantowski-Sachs cosmological model, one can thus transport directly the techniques developed for homogeneous but anisotropic LQC. One can refer to \cite{AB06,Vandersloot07,Chiou08} for detailed predications of the model, where it was shown by solving the effective dynamics that the black-hole singularity can be resolved as expected. Moreover, these treatments differ by how to choose the quantum parameters introduced to quantize the Hamiltonian constraint in detail. Roughly speaking, these analyses can be classified as the following types. The first is the so-called $\mu_o$-type, where the quantum parameters us chosen to be a global constant on the phase space \cite{AB06}.   The second one is the so-called $\bar\mu$-type\cite{Vandersloot07,Chiou08}, where, in contract with the $\mu_o$-type, the quantum parameters are chosen to be a function on the phase space. Although the key result of singularity resolution holds in both of the two approaches, the detailed consequent effective dynamics differ from one to the other. Several physically undesirable results are caused. For example, in some approaches, the quantum replacement of the classical singularity could occur in the low curvature region \cite{AB06,M06,CGP08}; while in other approaches the quantum corrections to the classical spacetime are large near the black hole horizon which is expected to be the classical regime \cite{Vandersloot07,Chiou08}. By considering these weaknesses, one recently proposed a new scheme \cite{AOS1,AOS2}, which will be referred as the AOS approach below. The AOS approach can be regarded as an average of the $\mu_o$-type and $\bar\mu$-type approaches, where the quantum parameters are required to be a Dirac observable, that is, a function on the phase space which are constant along each dynamics trajectory but may vary from one to another. The AOS approach provides a new effective description of the macroscopic Kruskal extension of Schwarzschild black holes incorporating corrections due to quantum geometry effects of loop quantum gravity. This effective dynamics not only resolves the central singularity in the Planck region but also keep the classical regime such as the horizon unchanged.

With these achievements, the AOS approach certainly should be generalized to the other black hole background to test its universality, which is the purpose of the present work. To achieve this goal, we choose the Janis-Newman-Winicour (JNW) spacetime\cite{JNW,JNW1,JNW2,JNW3} for investigation in the present work. The JNW spacetime is a solution to the Einstein gravity minimally coupled to a massless scalar field which on one hand is more complicated than the Schwarzschild solution and, on the other hand, can reduce to the Schwarzschild case when the scalar field vanishes. In the JNW spacetime, beyond the central singularity as that in the Schwarzschild case, there also exists another singularity, called the naked singularity, which will be replaced by the horizon in the Schwarzschild case. The $\mubar$ scheme of such system has been studied in \cite{Chiou08b}. We utilize the Hamiltonian framework in order to study the quantum effective dynamics. The Hamiltonian constraint of this model is an addition of the gravity part and the matter part, of which the gravity part is the same as that for the Schwarzschild case. Thus for the quantum effective dynamics, we can just transport the one given in \cite{AOS2,Chiou08b} to our model as the gravity part of the effective Hamiltonian constraint and leave the matter part as the classical one, just like the usual treatment in LQC \cite{Ash-view}. The effective Hamilton's equations then can be obtained by assuming the quantum parameters in the effective Hamiltonian are Dirac observables. The solutions to these equations can be written down by using the elliptic integral and thus properties of the solutions can be analyzed analytically and numerically.
According to the results,  the quantum dynamics well agrees with the classical one in the classical region and the quantum correction coming out in the quantum region resolves the singularities. These achievements do not sensitively depend on the value of the quantum parameters.

 To study the quantum effective dynamics more precisely, the quantum parameters must be fixed. As mentioned above, we want to generalize the AOS approach to the current model. However, a direct generalization of the AOS approach, which transports directly the key equations used to solve the parameters to our case, leads to a problem that there do not exist any real quantum parameters making the consequent equations hold. Therefore, the generalization is highly non-trivial.
To solve this problem, we come up with new equations by balancing the physical interpretation of the quantum parameters and some other physical consideration. The consequent results of the new proposed equations can recover the AOS scheme when the matter part vanishes. It should be mentioned here that some previous studies on loop quantization of the JNW spacetime with $\mubar$ scheme can be found in \cite{Chiou08b}. These previous studies reveal that the quantum evolution of the JNW spacetime will involve a large number of numerical calculations and thus becomes an ideal arena to highlight our approach.

This paper is organized as follows. In section \ref{sec:two}, the Hamiltonian framework of the classical theory of the JNW spacetime is presented. By considering the classical dynamics, we finally select the region with low curvature to set the initial data of the quantum effective dynamics. In section \ref{sec:three}, the quantum effective dynamics is discussed with using the assumption  that the quantum parameters are the Dirac observables. In section \ref{sec:four}, we present our approach of how to constraint the quantum parameters and discuss some consequences. Finally, the present work is summarized in section \ref{sec:con}.

\section{Classical theory}\label{sec:two}

\subsection{Preliminaries}

Consider a massless scalar field $\varphi$ minimally coupled to gravity with GR, the action reads
\begin{equation}
S[g_{ab},\varphi]=\int \dd^4 x\sqrt{-g}\left(\frac{1}{2\kappa}R-\frac{1}{2}g^{ab}\nabla_a\varphi\nabla_b\varphi\right)
\end{equation}
where $\kappa=8\pi G$. Canonical analysis leads to
\begin{equation}
S[g_{ab},\varphi]=\int \dd t\int\dd^3x \left(\frac{1}{\kappa}\dot{A}_a^iE_i^a+\dot{\varphi}\pi_\varphi-G_i\Lambda^i-C_aN^a-CN\right)
\end{equation}
where
\begin{equation}\label{eq:Hamilc}
\begin{aligned}
G_i=&\partial_bE^b_i+\epsilon_{ij}^{\ \ k}A_a^jE^a_k\\
C_a=&\frac{1}{\kappa\gamma}E_j^bF^j_{ab}+\pi_\varphi\partial_a\varphi\\
C=&\frac{1}{2\kappa}\frac{E^a_iE^b_j}{\sqrt{q}}\epsilon^{ij}_{\ \ k}\left(F^k_{ab}-(1+\gamma^2)\epsilon^k_{\ mn}K_a^mK^n_b\right)+\frac{1}{2}\frac{\pi_\varphi^2}{\sqrt{q}}+\frac{1}{2}\sqrt{q}q^{ab}\partial_a\varphi\partial_b\varphi
\end{aligned}
\end{equation}
are the Gaussian, diffeomorphism and Hamiltonian constraints respectively. Here we used the Ashtekar-Barbero variables where it should be noticed that $E_i^a(x)$ and $\pi_\varphi(x)$ are densitied fields of weight 1.

At the classical level, a spherically static solution to such a system was obtained by Janis, Newmann and Winicor(JNW)\cite{JNW,JNW1,JNW2,JNW3}.
The usual JNW metric is given by
\ba\label{eq:standardJNW}
\dd s^2=-\left(1-\frac{B}{r}\right)^\nu dt^2+\left(1-\frac{B}{r}\right)^{-\nu}dr^2+r^2\left(1-\frac{B}{r}\right)^{1-\nu}d\Omega^2
\ea
whereas the scalar field here reads \ba
\varphi=\frac{q}{B\sqrt{4\pi}}\ln\left(1-\frac{B}{r}\right)
\ea
with $q^2$ being the scalar
charge. The two parameters $B$ and $\nu$ are given by\ba
\nu=\frac{2m}{B}\\
B=2\sqrt{m^2+q^2}
\ea
with $m$ standing for the ADM mass. This solution recovers the Schwarzschild exterior spacetime by setting $\nu=1$. It should be noticed that in the JNW solution, $r$ cannot be smaller than $B$ and $r=B$ is a singularity for $\nu\neq 1$. However,  as inspired by the Schwarzschild interior metric, we can modified \eqref{eq:standardJNW} to be
\begin{equation}\label{eq:modifiedJNW}
\dd s^2=- \left(\frac{B}{\tau}-1\right)^{-\nu}\dd \tau^2+ \left(\frac{ B}{\tau}-1\right)^{\nu}\dd x^2+\left(\frac{B}{\tau}-1\right)^{1-\nu}\tau^2\dd\Omega^2
\end{equation}
with $\tau\in (0,B)$. It should be noticed that in \eqref{eq:modifiedJNW} the space and time coordinated exchange comparing with \eqref{eq:standardJNW}. The modified JNW metric \eqref{eq:modifiedJNW} gives us the Schwarzschild interior spacetime when the scalar field vanishes. It has two singularities, one locates at $r=B$ and the other is the center singularity at $r=0$. When it comes to the JNW spacetime below, we refer to the metric \eqref{eq:modifiedJNW}.

As in the interior of the Kruskal space-time, the homogeneous Cauchy slices $\Sigma$ of the JNW spacetime have topology $\mathbb{R}\times \mathbb{S}^2$. Thus we can introduce a fiducial metric $\mathring{q}_{ab}$ on $\Sigma$
\begin{equation}
\mathring{q}_{ab}\dd x^a\dd x^b=\dd x^2+r_o^2(\dd \theta^2+\sin^2\theta\dd\phi^2).
\end{equation}
where $x\in(-\infty,\infty)$, $\theta$ and $\phi$ are 2-sphere coordinates, and $r_o$ is a constant with dimensions of length. Since $\Sigma$ is non-compact in the direction $x$, we introduce an elementary  cell $\mathcal{C}\cong (0,L_0)\times \mathbb{S}^2$ in $\Sigma$ and restrict all integrals to this elemental cell to avoid the divergency problems of integrations.

Because of the symmetry, the scalar filed $\varphi$ is reduced to a constant and the fields $A_a^i(x)$, $E^a_i(x)$ and $\pi_\varphi(x)$ take the forms
\begin{equation}
\begin{aligned}
A_a^i\tau_i\dd x^a&=\frac{c}{L_0}\tau_3\dd x+b\tau_2\dd\theta-b\tau_1\sin\theta\dd\phi+\tau_3\cos\theta\dd\phi\\
E_i^a\tau^i\partial_a&=p_c\tau_3\sin\theta\partial_x+\frac{p_b}{L_0}\tau_2\sin\theta\partial_\theta-\frac{p_b}{L_0}\tau_1\partial_\phi\\
\pi_\varphi&=\frac{p_\varphi}{4\pi r_0^2L_0}\sqrt{\mathring{q}}
\end{aligned}
\end{equation}
The Poisson bracket of the phase space reads
\begin{equation}
\{c,p_c\}=2G\gamma,\ \{b,p_b\}=G\gamma,\ \{\varphi,p_\varphi\}=1.
\end{equation}
By employing the expression of connection and triad, the curvature $F$ of the connection $A$ and the extrinsic curvature $K$ take the forms
\ba
F_{ab}^i\tau_i\dd x^a\dd x^b&=&\frac{bc}{L_0}\tau_1\dd\theta\wedge \dd x+\frac{bc\sin\theta}{L_0}\tau_2\dd\phi\wedge\dd x+(\sin\theta-b^2\sin\theta)\tau_3\dd\phi\wedge\dd\theta\label{Fterm}\\
\gamma K_a^i\tau_i\dd x^a&=&\frac{c}{L_0}\tau_3\dd x+b\tau_2\dd\theta-b\tau_1\sin\theta\dd\phi\label{Kterm}
\ea
where $\gamma$ is the Barbero-immirzi parameter. The constraints except the Hamiltonian in \eqref{eq:Hamilc} vanish automatically. Taking into account of Eqs.(\ref{Fterm}) and (\ref{Kterm}), the Hamiltonian constraint in \eqref{eq:Hamilc} is reduced to
\begin{equation}
C=-\frac{1}{2\kappa}\frac{1}{|p_b|\sqrt{|p_c|}}\frac{2 p_b  \left(\left(b^2+\gamma ^2\right) p_b+2 b c p_c\right)}{\gamma ^2 }\frac{\sin(\theta )}{L_0}+\frac{1}{2}\frac{p_\varphi^2L_0}{16\pi^2 |p_b|\sqrt{p_c}}\sin\theta
\end{equation}
which becomes
\begin{equation}
\begin{aligned}
H:=\int_{\mathcal C} N C=-\frac{1}{2G\gamma}  \left(\left(b+\frac{\gamma ^2}{b}\right) p_b+2  c p_c\right)+\frac{\gamma p_\varphi^2L_0^2}{8 \pi p_b b}
\end{aligned}
\end{equation}
after smeared by a lapse function
$$N=\gamma\sgn(p_b)\sqrt{|p_c|}b^{-1}.$$

\subsection{The classical dynamics}

As the hamiltonian smeared with lapse function $N=\gamma\sgn(p_b)\sqrt{|p_c|}b^{-1}$ does not admit analytical solutions, we introduce a new lapse function $\tilde N$ in order to found  analytical solutions, which reads
\begin{equation}
\tilde{N}=8\pi G \gamma bp_b N.
\end{equation}
This lapse function leads to a new smeared Hamiltonian constraint
\begin{equation}
\tilde{H}:=\int_{\mathcal C}\tilde NC=8\pi G \gamma bp_b H.
\end{equation}
Let $\tilde{t}$ be the corresponding time to $\tilde{H}$. Noting that $p_c c$ is a constant of motion, we could define $p_c c=:L_0\gamma m$. Then the Hamiltonian constraint can be factorized as
\begin{equation}
-8\pi G \gamma ^3 p_b^2=8 \pi  G \gamma \left(b p_b\right)^2+16 \pi  G\gamma^2 m L_0 \left(b p_b\right)-2G^2 \gamma ^3 L_0^2 p_{\varphi }^2=:\kappa\gamma(bp_b-y_+)(bp_b-y_-),
\end{equation}
where
\begin{equation}
y_\pm=-\gamma L_0 m\pm \gamma L_0\sqrt{m^2+\frac{Gp_\varphi^2}{4\pi}}.
\end{equation}
The Hamilton's equation of $bp_b$ leads to
\begin{equation}
\frac{\dd (bp_b)}{\dd \tilde{t}}=\{bp_b,\tilde{H}\}=-8\pi G \gamma^3 p_b^2=\kappa\gamma(b p_b-y_+)(b p_b-y_-).
\end{equation}
which, with denoting $y=bp_b$,  can give us
\begin{equation}
\begin{aligned}
y(\tilde t)&=y_-+\frac{y_+-y_-}{1+e^{\kappa\gamma (y_+-y_-)\tilde{t}}}\\
p_b(\tilde{t})^2&=-\frac{1}{8\pi G \gamma^3}\frac{\dd y}{\dd\tilde{t}}=\frac{(y_+-y_-)^2}{4\gamma^2}\cosh\left(\frac{\kappa\gamma}{2}(y_+-y_-)\tilde{t}\right)^{-2}\\
b(\tilde{t})&=\frac{ \gamma }{  \left(y_+-y_-\right) }\left(y_+e^{-\kappa\gamma  \frac{(y_+-y_2)}{2}\tilde{t}}+y_-e^{\kappa  \gamma\frac{(y_+-y_2)}{2}\tilde{t}} \right)
\end{aligned}
\end{equation}
Then,
 $p_c(\tilde{t})$ and $c(\tilde{t})$ are easy to be obtained as
\begin{equation}
\begin{aligned}
p_c(\tilde{t})=p_c^{(0)}e^{2T(\tilde{t})}\\
c(\tilde{t})=c^{(0)} e^{-2T(\tilde{t})}
\end{aligned}
\end{equation}
where
\begin{equation}
T(\tilde{t})= \kappa \gamma y_+\tilde{t}-\ln \left(e^{\kappa\gamma  \left(y_+-y_-\right)\tilde{t}}+1\right).
\end{equation}
By substituting the above solutions to the spherical metric
\begin{equation*}
\dd s^2=-\tilde{N}^2\dd \tilde{t}^2+\frac{p_b^2}{|p_c|L_0^2}\dd x^2+|p_c|\dd\Omega^2,
\end{equation*}
we have
\begin{equation}\label{eq:metricnonstandard}
\begin{aligned}
\dd s^2=&-\gamma ^2|p_c^{(0)}|L_0^2\frac{ \left(\frac{\left(y_+-y_-\right)}{\gamma L_0\tau}-1\right){}^{\frac{y_-+y_+}{y_+-y_-}}}{\left(y_+-y_-\right){}^2}\dd \tau^2+\frac{\left(y_+-y_-\right)^2 \left(\frac{ \left(y_+-y_-\right)}{\gamma L_0\tau}-1\right){}^{\frac{y_-+y_+}{y_--y_+}}}{\gamma ^2|p_c^{(0)}|L_0^2}\dd x^2\\
&+\frac{|p_c^{(0)}|\gamma ^2 L_0^2}{\left(y_--y_+\right){}^2}\left(\frac{ \left(y_+-y_-\right)}{\gamma L_0\tau}-1\right){}^{\frac{y_-+y_+}{y_+-y_-}+1}\tau^2\dd\Omega^2
\end{aligned}
\end{equation}
where we defined
\begin{equation}
 \tau:=-\frac{ \left(y_+-y_-\right) \tanh \left(\frac{\kappa\gamma}{2} \left(y_+-y_-\right)\tilde{t}\right)}{2 \gamma L_0}+\frac{y_+-y_-}{2\gamma L_0}.
\end{equation}
By comparing Eq.\eqref{eq:metricnonstandard} with the JNW metric \eqref{eq:modifiedJNW}
\begin{equation}
\dd s^2=- \left(\frac{B}{\tau}-1\right)^{-\nu}\dd \tau^2+ \left(\frac{ B}{\tau}-1\right)^{\nu}\dd x^2+\left(\frac{B}{\tau}-1\right)^{1-\nu}\tau^2\dd\Omega^2
\end{equation}
we can fix the parameters as
\begin{equation}
B:=\frac{y_+-y_-}{\gamma L_0}= 2\sqrt{m^2+\frac{Gp_\varphi^2}{4\pi}},\ p_c^{(0)}=B^2,\ -\nu=\frac{y_-+y_+}{y_+-y_-}=-\frac{2m}{B}.
\end{equation}
The solutions to the equations of motion can be rewritten with the variables $B$ and $\nu$  as
\begin{equation}\label{eq:classicalsolution}
\begin{aligned}
b(\tilde{t})&=\frac{ \gamma}{2}\left((1-\nu)e^{-\kappa\gamma^2 L_0 \frac{B}{2}\tilde{t}}-(1+\nu)e^{\kappa\gamma^2 L_0  \frac{B}{2}\tilde{t}} \right)\\
p_b(\tilde{t})^2&=\frac{B^2L_0^2}{4}\cosh\left(\kappa\gamma^2L_0\frac{B}{2}\tilde{t}\right)^{-2}\\
p_c(\tilde{t})&=B^2 e^{2T(\tilde{t})}\\
c(\tilde{t})&=\frac{\nu\gamma L_0}{2B}e^{-2T(\tilde{t})}\\
T(\tilde{t})&= \frac{1}{2}\kappa \gamma^2L_0  B(1-\nu )\tilde{t}-\ln \left(e^{\kappa\gamma^2 L_0B\tilde{t}}+1\right).
\end{aligned}
\end{equation}
It is not difficult to verify that $T(\tilde{t})+$constant is actually the time with respect to the lapse function $N$. However, $T(\tilde{t})$ is not monotonous with respect to $\tilde{t}$ if $\nu\neq 1$. Its monotonicity changes at
\begin{equation}
\tilde{t}_0=\frac{1 }{ \kappa\gamma^2L_0B}\ln(\frac{1-\nu}{1+\nu}).
\end{equation}
More precisely, $T(\tilde{t})$ increases when $\tilde{t}\in (-\infty,\tilde{t}_0)$ but decreases when $\tilde{t}\in (\tilde{t}_0,\infty)$. This property of $T(\tilde{t})$ tells us that $T$ cannot be a globally defined time coordinate. In other words, a chart with $T$ as its time coordinate cannot cover the entire JNW spacetime. The entire spacetime is divided into two parts from $\tilde t=\tilde t_0$ by $T$. In both of the two part, $T$ ranges from $-\infty$ to $T(\tilde t_0)$, namely, $T(\tilde t)\in(-\infty, T(\tilde t_0))$ in both of the two parts. However, different physical pictures are presented as $T\to -\infty$ in the two parts.
For the part where
$\tilde t\in (-\infty,\tilde t_0)$, $T\to -\infty$ leads to the naked singularity, while for the other part, $T\to -\infty$ presents the central singularity. For the Schwarzschild's case where $\nu=1$, the naked singularity is replaced by the horizon. Then $T(\tilde t)$ becomes a globally defined time coordinate of the interior Schwarzschild spacetime.

At the moment $\tilde t=\tilde t_0$, $T(\tilde{t})$ takes its maximal value
\begin{equation}
T(\tilde t_0)=\frac{1}{2}(1-\nu) \ln \left(\frac{1-\nu}{1+\nu}\right)-\ln \left(\frac{2}{1+\nu}\right).
\end{equation}
and $b(\tilde t_0)$ $c(\tilde t_0)$, $p_b(\tilde t_0)$ and  $p_c(\tilde t_0)$ read
\begin{equation}\label{eq:initialdata}
\begin{aligned}
b(\tilde t_0)&=0\\
c(\tilde t_0)&=\frac{2 \gamma L_0}{B}\frac{  \nu }{1- \nu ^2} \left(\frac{1-\nu}{1+\nu}\right)^{\nu }\\
p_b(\tilde t_0)&=\frac{1}{2} B L_0 \sqrt{1-\nu ^2}\\
p_c(\tilde t_0)&=\frac{1}{4} B^2 \left(1-\nu ^2\right) \left(\frac{1-\nu}{1+\nu }\right)^{-\nu }.
\end{aligned}
\end{equation}
The scalar curvature is
\begin{equation}\label{eq:classicalR}
R=-\frac{8 }{B^2 \left(1-\nu ^2\right)}\left(\frac{1-\nu }{1+\nu}\right)^{\nu }.
\end{equation}
which will be much smaller than $ 1$ for $B\gg 1$ and all of $0<\nu<1$. Thus around $\tilde t=\tilde t_0$ is a classically region. In other words, when $\tilde t=\tilde t_0$, the quantum correction should be tiny such that the quantum dynamics is compatible with the classical one. This guides us to initialize the quantum dynamics.

\section{Quantum theory}\label{sec:three}
We adopt the same convention in \cite{AOS1,AOS2}. Then the effective Hamiltonian of the JNW spacetime reads
\ba
H_{\rm eff}=-\frac{1}{2G\gamma}\left[2\frac{\sin(\delta_c c)}{\delta_c}p_c+\left(\frac{\sin(\delta_b b)}{\delta_b}+\frac{\gamma^2\delta_b}{\sin(\delta_b b)}\right)p_b\right]+\frac{\gamma \delta_bp^2_\varphi L_0^2}{8\pi \sin(\delta_b b)p_b}
\ea
with
\begin{equation}
N=\gamma\sgn(p_b)\sqrt{|p_c|}\frac{\delta_b}{\sin(\delta_bb)}
\end{equation}

In the quantum theory, we choose the time coordinate with respect to same $N$, which corresponds to $T(\tilde t)$ of the classical theory define in \eqref{eq:classicalsolution}. This time coordinate in the quantum dynamics will be also denoted as $T$. Taking advantage of the choice of this time coordinate, we can solve the evolution of $p_c$ and $\tan(\delta_c c)$ analytically. By using the Hamilton's equations for $p_c$ and $ c$
\ba
\frac{\dd p_c}{\dd T}&=&2p_c\cos(\delta_c c), \nn\\
\frac{\dd c}{\dd T}&=&-2\frac{\sin(\delta_c c)}{\delta_c},\nn\\
\ea
we have
\ba
\tan(\frac{\delta_c c(T)}{2})&=&c^{(0)}e^{-2T}\\
p_c(T)&=&\pm\frac{\zeta}{2}  \left(\frac{e^{2 T}}{c_0}+c^{(0)} e^{-2 T}\right)
\ea
with $\zeta=p_c\sin(\delta_c c)$ being the constant of motion corresponding to $cp_c$ in the classical theory \footnote{It is noticed here that, we have assumed $\delta_c$ and $\delta_b$ are Dirac observables. The validity of such assumption can be confirmed by using the same strategy adapted in the Appendix A of \cite{AOS2}.}.  As discussed under \eqref{eq:initialdata}, we should set the initial data such that the quantum dynamics around $T=0$ is compatible with the classical one around $\tilde t=\tilde t_0$. This principle leads to
\begin{equation}
c^{(0)}=\frac{\gamma L_0\delta_c}{B}\frac{\nu}{1-\nu^2}\left(\frac{1-\nu}{1+\nu}\right)^\nu.
\end{equation}
and
\begin{equation}
\zeta=\delta_cL_0\gamma \frac{\nu B}{2}.
\end{equation}
Thus we have
\begin{equation}\label{eq:solutionpc}
\begin{aligned}
\tan  (\frac{\delta_c c}{2})&=\frac{\gamma  L_0 \delta _c}{B}\frac{\nu }{1-\nu ^2}\left(\frac{1-\nu}{1+\nu }\right)^{\nu }e^{-2 T}\\
p_c&=\frac{B^2}{4}(1-\nu ^2)\left(\frac{1-\nu}{1+\nu }\right)^{-\nu }\left(e^{2 T}+\frac{\left(\gamma  L_0 \delta _c\right)^2}{B^2}\frac{\nu ^2}{\left(1-\nu ^2\right)^2}\left(\frac{1-\nu}{1+\nu }\right)^{2 \nu }e^{-2 T}\right)
\end{aligned}
\end{equation}
which recovers the classical solution \eqref{eq:classicalsolution} if $\tan(\delta_cc/2)\ll 1$.

Let us denote $\cos(\delta_b b)=:\xi$ below for abbreviation. The Hamilton's equation for $\xi$ gives
\begin{equation}\label{eq:xidot}
\begin{aligned}
\frac{\dd \xi}{\dd T}&=\frac{1}{2} (1-\xi^2+\gamma^2\delta_b^2)+\frac{G\gamma ^2  \delta _b^2 p_{\varphi }^2L_0^2}{8\pi  p_b^2}
\end{aligned}
\end{equation}
By employing the constraint equation $H_{\rm eff}=0$, we get
\begin{equation}\label{eq:pbofxi}
\begin{aligned}
p_b^{(\pm)}
=\frac{ L_0m\gamma \delta_b \left(-\sgn(\sin(\delta_bb)) \nu\sqrt{1 -  \xi ^2}\pm \sqrt{(1-\nu^2) \gamma ^2 \delta_b^2+  \left(1-\xi ^2\right)}\right)}{ \nu\left(1-\xi^2+\gamma ^2 \delta_b^2\right)}.
\end{aligned}
\end{equation}
Substituting \eqref{eq:pbofxi} into \eqref{eq:xidot}, we have that
\begin{itemize}
\item for $\sgn(\sin(\delta_b b))>0$, the two solutions  $p_b^{(\pm)}$ read
\begin{equation}\label{eq:detail1}
\begin{aligned}
\frac{\dd T}{\dd \xi}
=\frac{1}{\gamma ^2 \delta_b^2-\xi ^2+1}\mp \frac{\nu(1 -  \xi ^2)}{\left(\gamma ^2 \delta_b^2-\xi ^2+1\right) \sqrt{(1-\nu^2)\gamma ^2 \delta_b^2+\left(1-\xi ^2\right)}\sqrt{1-\xi^2}}=:R_1(\xi)\mp R_2(\xi).
\end{aligned}
\end{equation}
\item for $\sgn(\sin(\delta_b b))<0$, the two solutions  $p_b^{(\pm)}$ read
\begin{equation}\label{eq:detail2}
\begin{aligned}
\frac{\dd T}{\dd \xi}
=\frac{1}{\gamma ^2 \delta_b^2-\xi ^2+1}\pm \frac{\nu(1 -  \xi ^2)}{\left(\gamma ^2 \delta_b^2-\xi ^2+1\right) \sqrt{(1-\nu^2)\gamma ^2 \delta_b^2+\left(1-\xi ^2\right)}\sqrt{1-\xi^2}}=:R_1(\xi)\pm R_2(\xi).
\end{aligned}
\end{equation}
\end{itemize}
Then
without loss of generality, we will define the functions $T^\pm (\xi)$ as solutions to
\begin{equation}\label{eq:dTpmdxi}
\frac{\dd T^\pm}{\dd \xi}=R_1(\xi)\mp R_2(\xi).
\end{equation}
More precisely, we have
\begin{equation}
\begin{aligned}
T^\pm(\xi)=\frac{-1}{2\sqrt{1+\gamma^2\delta_b^2}}\ln\left(\frac{\sqrt{1+\gamma^2\delta_b^2}-\xi}{\sqrt{1+\gamma^2\delta_b^2}+\xi}\,\frac{\sqrt{1+\gamma^2\delta_b^2}+1}{\sqrt{1+\gamma^2\delta_b^2}-1}\right)\mp \int_1^\xi R_2(\xi')\dd \xi'
\end{aligned}
\end{equation}
where the initial condition $T(\xi=1)=0$ is considered because we want the quantum dynamics around $T=0$ to recover the classical one around $\tilde t=\tilde t_0$.  Then all information of the quantum effective dynamics is stored in the two functions $T^\pm(\xi)$.

The integration of $R_2$ above is actually a summation of the standard elliptic integrals, that is
\begin{equation}
\begin{aligned}
\int_1^\xi R_2(x)\dd x
=k\nu \Big(F(\arcsin(\xi)\,|\,k)-F(\pi/2\,|\,k)\Big)-k\nu(1+n)\Big(\Pi(n,\arcsin(\xi)\,|\,k)-\Pi(n,\pi/2,k)\Big)
\end{aligned}
\end{equation}
where $F(\theta\,|\,k)$ and $\Pi(n,\theta\,|\,k)$ are the first and the third kind of elliptic integrals (refer to Appendix \ref{app:ellip} for details) and the parameters $k$ and $n$ are defined as
\begin{equation}\label{eq:kn}
k^2=\frac{1}{1+(1-\nu^2)\gamma^2\delta_b^2},\ n=-\frac{1}{1+\gamma^2\delta_b^2}.
\end{equation}
Therefore, $T^\pm(\xi)$ takes the form
\begin{equation}\label{eq:Tofxisol}
T^\pm(\xi)=\frac{-1}{2\sqrt{1+\gamma^2\delta_b^2}}\ln\left(\frac{\sqrt{1+\gamma^2\delta_b^2}-\xi}{\sqrt{1+\gamma^2\delta_b^2}+\xi}\right)\mp k\nu F(\arcsin(\xi)\,|k)\pm k\nu(1+n)\Pi(n,\arcsin(\xi)|k)+T_0^\pm
\end{equation}
with
\begin{equation}\label{eq:initialT}
T_0^\pm=\frac{-1}{2\sqrt{1+\gamma^2\delta_b^2}}\ln\left(\frac{\sqrt{1+\gamma^2\delta_b^2}+1}{\sqrt{1+\gamma^2\delta_b^2}-1}\right)\pm k\nu F\left(\pi/2|k\right)\mp k\nu(1+n)\Pi(n,\pi/2|k).
\end{equation}
A numerical method to compute the elliptic integrals, and therefore to compute $T^\pm(\xi)$, is introduced in appendix \ref{app:ellip}.

\begin{figure}
\centering
\includegraphics[width=0.5\textwidth]{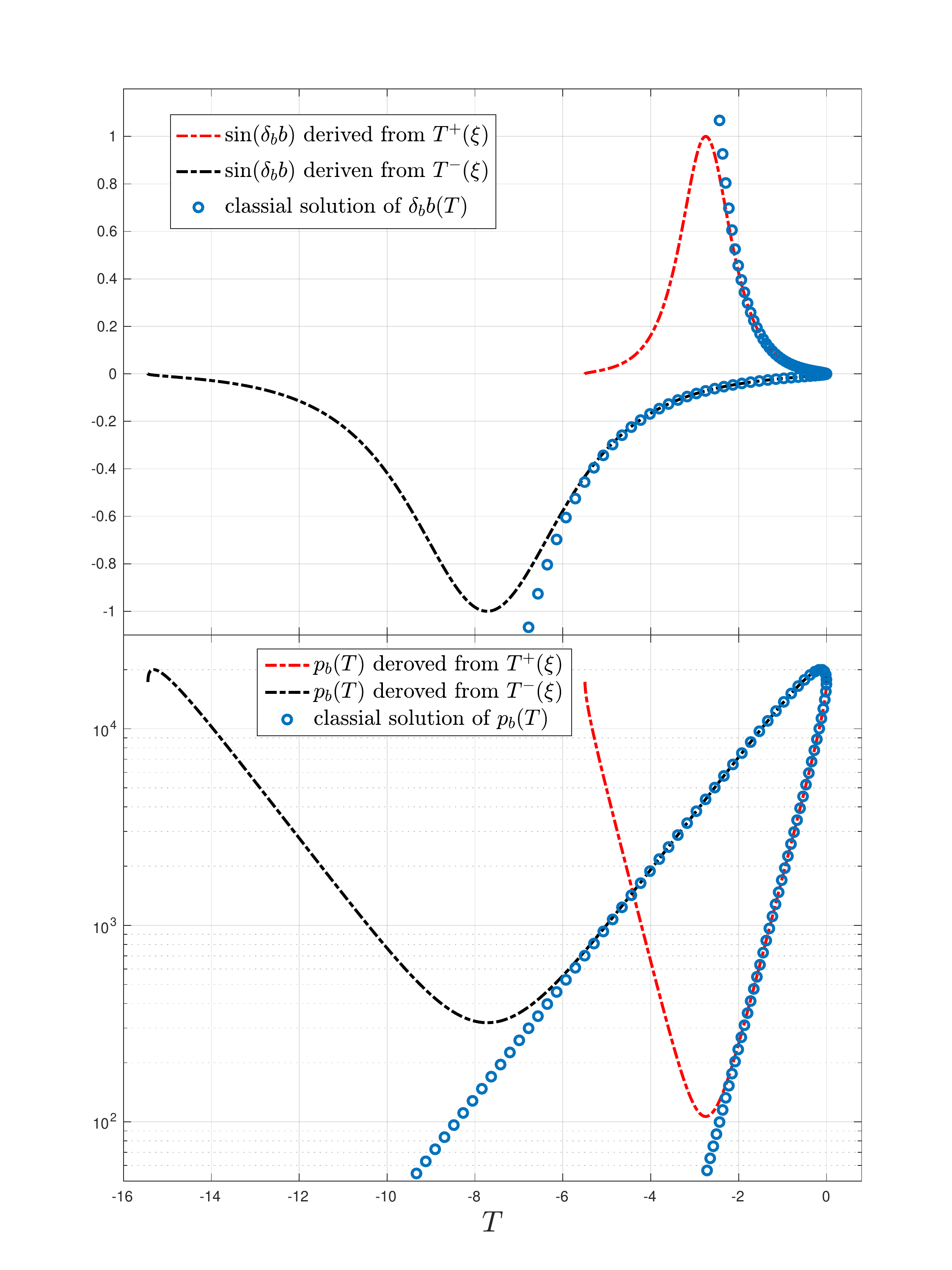}
\caption{Plots of the evolution of $\sin(\delta_b b)$ and $p_b$ as functions of $T$ derived from $T^\pm(\xi)$. As a comparison, the classical results of $\delta_b b(T)$ and $p_b(T)$ derived from \eqref{eq:classicalsolution} are also shown. The parameters are chosen to be
 $m=10^4$, $\nu=0.5$, $G=1=\hbar$, $\gamma=0.2375$, $L_0=1$ and $\delta_b=0.0340$ which is computed by \eqref{eq:deltab}. By comparing the quantum dynamics with the classical one, we conclude that $T^-(\xi)$, corresponding to $p_b^{(-)}$ in \eqref{eq:pbofxi}, resolves the central singularity and that $T^+(\xi)$, corresponds to the naked singularity side. }\label{fig:realsol}
\end{figure}

Now we have two solutions $T^\pm(\xi)$ to the equation \eqref{eq:xidot}. The meaning of these two solutions is understood by the following discussion.
In the classical theory, the spacetime must be divided at least into two parts if $T(\tilde t)$ is chosen as a time coordinate. In one part $b\in (-\infty, 0)$ and in the other part $b\in (0,\infty)$. Therefore, the spacetime constructed from the quantum effective dynamics is expected to be a union of two sub-manifolds. In each sub-manifold, $T$ is a well defined time coordinate. However, $T$, as emphasized, cannot be defined globally on the entire spacetime. Additionally, in the sub-manifold corresponding to the classical one where $b\in  (-\infty, 0)$, $\sin(\delta_b b)$ is expected to be $\sin(\delta_b b)\in (-1,0)$, and in the other, $\sin(\delta_b b)\in (0,1)$. In other words, the sign of $\sin(\delta_b b)$ changes if we move from one sub-manifold to the other. Thus, according to \eqref{eq:pbofxi}, in order to keep the continuity of $p_b$ at the moment $T=0$, namely in the Cauchy surface of the spacetime where $\xi=1$ and where the two sub-manifold is glued, we must assign, saying, $p_b^{(+)}$ to both of the two sub-manifold. This leads to the fact that in one sub-manifold solution to \eqref{eq:xidot} is given by $T^+(\xi)$ while in the other, it is given by $T^-(\xi)$ (referring to \eqref{eq:detail1} and \eqref{eq:detail2}).  According to the above discussion, figure \ref{fig:realsol} gives a plot of the evolution of $\sin(\delta_b b)$ and $p_b$ as functions of $T$ derived from $T^\pm(\xi)$, where, as a comparison, the classical results of $\delta_b b(T)$ and $p_b(T)$ derived from \eqref{eq:classicalsolution} are also shown. According to the numerical results shown in the figure, the difference between the classical and quantum dynamics increases gradually as $b$ becomes larger and larger, and simultaneously $p_b$ becomes smaller and smaller. The decreases of $p_b$ in the two parts of the entire spacetime driven by the classical dynamics are both replaced in the quantum effective dynamics by a bounce, which indicates to us how the two singularities in the JNW spacetime is resolved. Additionally, by comparing with the classical dynamics, it can be easily concluded that $T^-(\xi)$, corresponding to $p_b^{(-)}$ in \eqref{eq:pbofxi}, resolves the central singularity and that $T^+(\xi)$, correspond to the naked singularity side.

Finally, as $\xi\in [-1,1]$, the time coordinate $T$ ranges in a finite interval, namely $T^\pm(-1)\leq T\leq 0$ in the two charts respectively. It should be noticed that the consequent spacetime is geodesic incomplete. While how to extend this spacetime will be left as our further topic.

\section{The quantum parameters of $\delta_b$ and $\delta_c$}\label{sec:four}
As it is known, the quantum parameters $\delta_b$ and $\delta_c$ are introduced to regularize the curvature $F^i_{ab}(x)$, namely to express the $F_{ab}^i$ with holonomies \cite{Ash-view,AOS1}. More precisely,  taking $F_{\theta,\phi}^i\tau_i$ for instance, it is regularized as the following:
we first evaluate the ratio $(h_{\square(\theta,\phi)}-1)/\left({\rm Ar}(\square(\theta,\phi))\right)$ with $h_{\square(\theta,\phi)}$ being the holonomy around a closed rectangular plaquette $\square(\theta,\phi)$ within the $\theta$-$\phi$ 2-sphere enclosing an area ${\rm Ar}(\square(\theta,\phi))$, and then take the limit ${\rm Ar}(\square(\theta,\phi))\to 0$. Then by replacing the holonomy as an operator, the quantum operator $\hat F_{\theta\phi}^i\tau_i$  is given. However, by considering the underlying quantum geometry,  when defining the operator, we take the limit as $\square(\theta,\phi)$ shrinks to $\Delta$ with $\Delta$ being the area gap proposed by LQG\cite{Ash-view,AOS1}. In the same manner, operators corresponding to the other two non-vanished curvature components $F_{\phi,x}^i\tau_i$ and $F_{\theta,x}^i\tau_i$ are defined  by using holonomies along plaquette $\square(\phi,x)$ and $\square(\theta,x)$ in the $\phi-x$ and $\theta-x$ planes respectively. Finally we have
$$\hat F_{ab}^i\tau_i=\lim_{{\rm Ar}(\square(a,b))\to \Delta}\frac{1}{{\rm Ar}(\square(a,b))}(h_{\square_{ab}}-1)$$
where $\square_{ab}$ lies in the $a-b$ plane. In this procedure, the quantum parameter $\delta_b$ and $\delta_c$ are introduced, where $\delta_b$ is the ratio length of each link in the plaquette $\square(\theta,\phi)$  and $\delta_c$, the ratio length of the links along the $x$-direction in the plaquettes $\square(x,\theta)$ and $\square(x,\phi)$ in the fiducial cell $\mathcal C$.

Roughly speaking there are two strategies to choose the quantum parameters $\delta_b$ and $\delta_c$. The first one is the so-called $\mu_o$-type strategy, where the quantum parameters are chosen to be a global constant on the phase space such that the areas of the plaquettes measured with the fiducial metrics $\mathring{q}$ limit to $\Delta$, while, in contract with the $\mu_o$-type strategy, the second one, the so-called $\bar\mu$-type strategy, choose the quantum parameters to be a function on the phase space so that the areas measured with the physical metrics $q_{ab}$ limit to $\Delta$. In the Schwarzschild case, both of the two strategies are considered \cite{AB06,Vandersloot07,Chiou08}. However it is revealed that the consequent quantum dynamics has some physically undesirable results. To overcome the limitations, a new scheme  straddling between the $\mu_o$ and $\bar \mu$ schemes is introduced by \cite{AOS1,AOS2}. The idea is to define $\delta_b$ and $\delta_c$ as Dirac observables which keep constant along the quantum effective trajectories but may differ from one trajectory to another. This idea will be also adopted in the current paper.

The strategy in \cite{AOS1} to fix the Dirac observables $\delta_b$ and $\delta_c$ is to demand that the plaquette $\square(\theta,\phi)$ and $\square(\phi,x)$ enclose $\Delta$ physically when evaluated on the transition surface $\mathcal T$ with $\mathcal{T}$ being the space-like, 3-dimensional surface which replaces the classical singularity and glues the trapped and an anti trapped regions. By considering the physical meaning of the quantum parameters $\delta_b$ and $\delta_c$, the authors of  \cite{AOS1,AOS2} selected them by demanding that the two plaquettes within the $\theta-\phi$ 2-shperes and $\phi-x$ planes,  used to regularize the curvature $F_{ab}^i$, enclose the minimum area when evaluated in $\mathcal{T}$. However, we will see below that this requirement cannot be satisfied always for $\nu \neq 1$.

Let us denote $\tbbpm$ as the moments when $p_b$ bounces, where the index $+$ corresponds to the bounce of $p_b$ resolving the naked singularity and the index $-$ represents the bounce resolving the central singularities respectively. Then according to \eqref{eq:pbofxi}, the area of $\square(\phi,x)$ evaluated in $\tbbpm$ is
\begin{equation}\label{eq:arpx}
\arpx^\pm=2\pi \delta_b\delta_c|p_b|\Big|_{T=\tbbpm}=2\pi \delta_b\delta_cL_0\frac{ B \gamma \delta_b( \sqrt{(1-\nu^2)\gamma^2\delta_b^2+1}\mp \nu)}{2(1-\gamma^2\delta_b^2)}\cong ( 1\mp \nu)\pi \gamma\delta_b^2\delta_cL_0B.
\end{equation}
 Similarly, let us denote $\tbc$ as the moment when $p_c$ bounces. The area of $\square(\theta,\phi)$ evaluated in $\tbc$ is
\begin{equation}\label{eq:artp}
\artp=\left.4\pi \delta_b^2p_c\right|_{T=\tbc}=2\nu\pi \gamma\delta_b^2\delta_c L_0 B
\end{equation}
where \eqref{eq:solutionpc} has been used. As a consequence of \eqref{eq:arpx} and \eqref{eq:artp}, for $\nu=1$, the areas $\arpx^-$ and $\artp$ can equal the area gap $\Delta$ simultaneously if and only if $\tbb=\tbc$ \footnote{Since the naked singularity disappears classically in this case, the value of $\arpx^+$ corresponding to the factor $1-\nu$ in \eqref{eq:arpx} disappears too, that is, $\arpx$ takes only $\arpx\Big|_{T=\tbb}\cong 2\pi\gamma \delta_b^2\delta_c L_0 B$. }. Actually that both $\arpx^-$ and $\artp$ equal $\Delta$ is the requirement used in \cite{AOS2} to determine the quantum parameters. However, this requirement is invalid for the case when $\nu<1$. This can also be seen from the numerical calculation presented in the Appendix \ref{app:nogo}, where we show that the equations
\begin{equation}\label{eq:Aeq}
\begin{aligned}
2\pi \delta_b\delta_c|p_b|\Big|_{T=\tbc}&=\Delta\\
4\pi \delta_b^2p_c\Big|_{T=\tbc}&=\Delta,
\end{aligned}
\end{equation}
 will have no solution for $\nu$ smaller than 1. It should be remarked that in this equation, $p_b$ takes the value
 \begin{equation}
 |p_b|=\frac{ L_0m\gamma \delta_b \left(\nu\sqrt{1 -  \xi ^2}+ \sqrt{(1-\nu^2) \gamma ^2 \delta_b^2+  \left(1-\xi ^2\right)}\right)}{ \nu\left(1-\xi^2+\gamma ^2 \delta_b^2\right)}.
 \end{equation}
 with which Eq.(\ref{eq:Aeq}) can return to the equations introduced in \cite{AOS2} for $\nu=1$.

 The result that \eqref{eq:Aeq} has no solution is regarded to be physically desirable. Otherwise, the Ricci scalar curvature at $T=\tbc$, which is
\begin{equation*}
R\Big|_{T=\tbc}\cong\frac{\pi \left(\nu +1\right)^2 }{2\nu ^4\gamma^2}\frac{  \left(8{\rm Ar}(\square(\phi,x)) ^2 {\rm Ar}(\square(\phi,\theta)) \nu ^2+{\rm Ar}(\square(\phi,\theta))^3\right)}{ {\rm Ar}(\square(\phi,x))^4  }+o(\delta_b^2),
\end{equation*}
will increase to $\infty$ as $\nu$ decreases to $0$, which is contradictory to the
results indicated by LQC and the studies on loop quantized Schwarzschild interior spacetime, that loop quantum effect could lead to an upper bound to the scalar curvature \cite{Ash-view,AOS2}.

Now the problem on how to constraint the quantum parameters in the current model comes out. At first, because of the physical meaning of $\delta_b$ and $\delta_c$, the existence of the area gap proposed by LQG should be considered. However, it should be noticed that the existence of the area gap does not mean that any area in a spacetime cannot be smaller than $\Delta$. A typical counterexample is a null plaquette whose area always vanishes. This is actually the case for the plaquette  $\square(\phi,x)$ in the Schwarzschild spacetime when evaluated in the horizon. The situation is similar for the JNW spacetime considered in the current paper. When $\nu\cong 1$, in the surface $T=\tbbm$ where the bounce of $p_b$ to replace the naked singularity occurs, the metric will be almost degenerate along the $x$ direction and well behaved along the $\theta$ and $\phi$ directions. That is, the surface $T=\tbbm$ almost behaves as a horizon. Therefore, considering the limitation that the area of $\square(\phi,x)$ cannot be smaller than $\Delta$ when referring to the moment $T=\tbbm$ seems not to be that physically meaningful. Moreover, if we keep that $\arpx^-=\Delta$, it will lead to that $\artp<\Delta$ which is unphysical. Therefore in the current paper, we choose the equation $\artp=\Delta$ as one condition to determine the quantum parameters by considering the existence of the are gap.

Moreover, there have two quantum parameters $\delta_b$ and $\delta_c$, so we need two equations to completely fix them. The second equation we proposed is
\begin{equation}\label{eq:tbEqtc}
\tbbp=\tbc.
\end{equation}
which, as mentioned above, is automatically satisfied for the Schwarzschild case if one used the strategy in \cite{AOS1,AOS2}. Moreover, the equation \eqref{eq:tbEqtc} also means that the volume of the fiducial cell reaches its minimal in $T=\tbbm$ and, thus the energy density and the scalar curvature increase to their maximal at this moment. In summary, the quantum parameters will be determined by the equations
\begin{equation}\label{eq:mainequations}
\begin{aligned}
\tbbp&=\tbc\\
4\pi \delta_b^2p_c\Big|_{\tbc}&=\Delta.
\end{aligned}
\end{equation}

Now it comes to how to solve the proposed equations.
By employing the exact solutions of $c$ and $p_c$, we will obtain that
\begin{equation}\label{eq:tbc}
\tbc=\frac{1}{2} \ln \left( \left(\frac{1-\nu}{1+\nu }\right)^{\nu }\frac{  \nu }{1- \nu ^2}\frac{\gamma L_0 \delta_c}{B}\right).
\end{equation}
In addition, Eq.(\ref{eq:Tofxisol}) gives
\begin{equation}\label{eq:tbbpval}
\tbbp=\int_1^0 \frac{\dd T}{\dd \xi}\dd\xi=\frac{-1}{2}\sqrt{-n}\ln\left(\frac{\sqrt{-n}+1}{1-\sqrt{-n}}\right)+k\nu nR_F(1,1-k^2,0)-
\frac{k\nu n(1+n)}{3}R_J(1,1-k^2,0,1+n).
\end{equation}
In the equations \eqref{eq:tbc} and \eqref{eq:tbbpval}, we used the fact that
\begin{equation*}
\cos(\delta_c c)\Big|_{T=\tbc}=0=\cos(\delta_b b)\Big|_{T=\tbbp}.
\end{equation*}
The elliptic integrals in \eqref{eq:tbbpval} have been rewritten by using the Carlson's form \eqref{eq:carlsonform} whose relation with the Legendre forms is given in
\eqref{eq:relationce}. By using the assumption that $\delta_b\ll 1$ which leads to $1-k^2\ll1$ and $1+n\ll 1$, the asymptotic behavior of $\tbb$ can be derived by using the formula \eqref{eq:assRF} and \eqref{eq:assRJ}. The result is
\begin{equation}\label{eq:assTbb}
\begin{aligned}
\tbbp\cong\frac{-1}{2}\sqrt{-n}\ln\left(\frac{\sqrt{-n}+1}{1-\sqrt{-n}}\right)+k\nu n\ln\left(\frac{4}{\sqrt{1-k^2}}\right)-k\nu n\sqrt{\frac{n+1}{k^2+n}}\ln\left(\frac{\sqrt{(k^2+n)}+\sqrt{n+1}}{\sqrt{1-k^2}}\right)
\end{aligned}
\end{equation}
which leads to
\begin{equation}\label{eq:tbbformula}
\begin{aligned}
e^{\tbbp}&\cong\frac{1}{2^{2 \nu +1}} \frac{\sqrt{1+\nu}^{\,\nu +1}}{ \sqrt{1-\nu }^{\,1-\nu} }(\gamma\delta_b)^{1+\nu }.
\end{aligned}
\end{equation}
Then the equation \eqref{eq:tbEqtc} becomes
\begin{equation}\label{eq:deltac_deltab}
\begin{aligned}
\delta_c=\frac{B}{\gamma L_0}\frac{1- \nu ^2} {  \nu } \left(\frac{1-\nu}{1+\nu }\right)^{-\nu }e^{2\tbbp}\cong \frac{B}{\gamma L_0}\frac{1} {  \nu } \frac{1}{4^{\,2 \nu +1}} (\nu +1)^{\,2\nu +2}(\gamma\delta_b)^{\, 2\nu+2}
\end{aligned}
\end{equation}
Substituting this result \eqref{eq:deltac_deltab} into the second equation of \eqref{eq:mainequations} which more precisely is
\begin{equation}
\artp=2\nu\pi\gamma\delta_b^2\delta_c L_0 B=\Delta
\end{equation}
we finally have that
\begin{equation}\label{eq:deltab}
\delta_b\cong \left(\frac{\sqrt{\Delta}  } {\sqrt{2\pi} \left(\frac{\nu +1}{2}\gamma\right)^{ \nu +1}\frac{B}{2^\nu} }\right)^{\frac{1}{ \nu +2}}.
\end{equation}
Meanwhile, $\delta_c$ takes
\begin{equation}\label{eq:deltac}
\begin{aligned}
  \delta_cL_0\cong \frac{1} {  \nu } \left(\left(\frac{(1+\nu)^2\Delta}{8\pi}\right)^{\frac{1+\nu}{\nu}} \frac{\gamma}{ 4 B}\right)^{\frac{\nu}{\nu+2}}
 \end{aligned}
\end{equation}

Several remarkable features of our results should be discussed. First,
these results of $\delta_b$ and $\delta_c$ will recover that in the Schwarzschild case given in \cite{AOS1,AOS2} for $\nu=1$. Moreover, for the area ${\rm Ar}(\square(\phi,x))$ of the plaquette $\square(\phi,x)$, we obtain that
\begin{equation}
\begin{aligned}
{\rm Ar}(\square(\phi,x))&=2\pi \delta_b\delta_c|p_b|\Big|_{\mathcal T}=   \frac{\nu+1}{2\nu}  \Delta\geq \Delta.
\end{aligned}
\end{equation}
where the equality can be achieved only for $\nu=1$.

Secondly, the equation \eqref{eq:deltac} tells us that $\delta_c L_0$ is inversely proportional to  $\nu$, which means that $\delta_c$ can be very large for small $\nu$ if $L_0$ is fixed. This is incompatible  with the usual assumption that the quantum parameters take small values so that the quantum dynamics can be compatible with the classical ones in the classical regions.
However, in the current work such an assumption for $\delta_c$ is actually not necessary.
First of all, in the above derivation, we only assume that $\delta_b$ takes small value, which is coincide with the finally result \eqref{eq:deltab}.
Secondly, because the $x$-direction is the non-compact direction of the Cauchy surface, the link along the $x$-direction in the plaquettes $\square(x,\theta)$ and $\square(x,\phi)$ can be chosen with an arbitrary length. Therefore, $\delta_c$ can be arbitrarily large, with recalling that $\delta_c L_0$ is the length of our link with respect to the fiducial metric $\mathring{q}_{ab}$.
Moreover, the quantum dynamics actually can recover the classical one in the classical regions even though the parameter $\delta_c$ is large. Substituting the value of $\delta_c$ in \eqref{eq:deltac} into \eqref{eq:solutionpc}, we get that $\tan(\delta_c c/2)\ll 1$ around $T=0$.
Then in the classical region where $T\ll 1$,
 $$\frac{\delta_c c}{2}=\arctan\left( \tan(\frac{\delta_c c}{2})\right)\cong  \tan(\frac{\delta_c c}{2}),$$
  which gives a coincide result $c(T)$ with the classical solution $c(\tilde t)$ given in \eqref{eq:classicalsolution}. It should be reminded here that the time coordinate $T$ used in the quantum theory is the same as the function $T(\tilde t)$ up to a constant. In addition, the coincidence for $p_c$ can be verified easily by using the condition that $\delta_c L_0 \nu/B \ll 1$. Finally, since $\delta_b$ itself is small, the quantum and classical dynamics for $\delta_b$ and $p_b$ are compatible classically, which actually has been shown in Fig.  \ref{fig:realsol} as an example. Therefore, as a conclusion, large values of $\delta_c$ for small $\nu$ is not a problem in the sense that it will not lead to any incompatibility between the quantum and classical dynamics classically.

\begin{figure}
\centering
\includegraphics[width=0.6\textwidth]{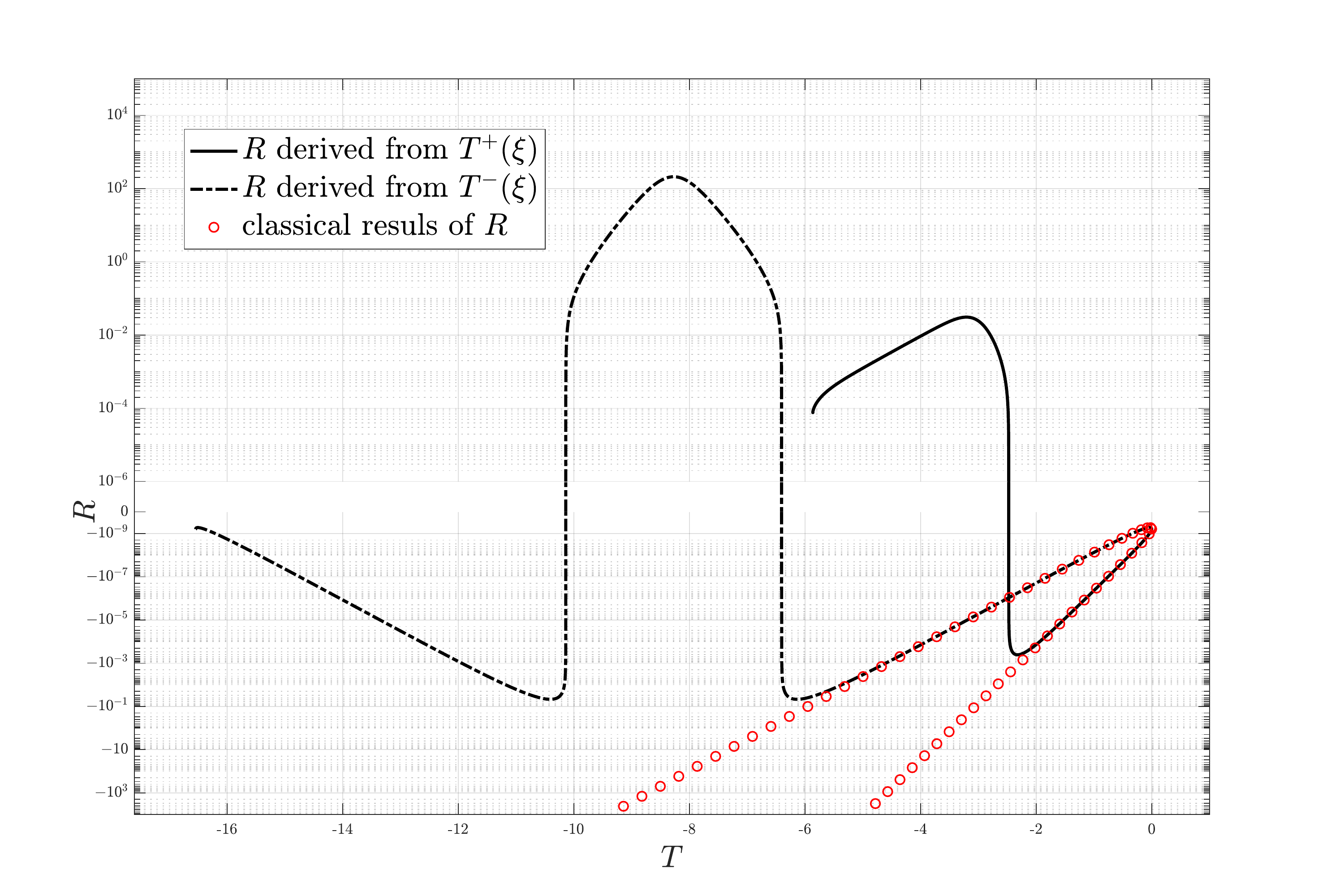}
\caption{Numerical results of the evolution of the scalar curvature $R$, for both of the classical and the quantum dynamics with $B=10^7$.  According to the result, the classical divergences as a consequence of the classical singularities is replaced by bouncing evolution of $R$. The quantum dynamics changes the sign of the curvature and the classical singularities with negatively infinity curvature is finally replaced by positive-valued curvature bounces. The parameter in this plot is chosen as $L_0=1=G=\hbar$, $\gamma=0.2375$, $\nu=0.5$ and $B=10^5$}\label{fig:riemanncurvature}
\end{figure}

Finally, let us consider the evolution of the scalar curvature $R$. By definition of $R$ and the Hamilton's equation, the scalar curvature can be expressed as
\begin{equation}\label{eq:curvature}
\begin{aligned}
R=&\frac{1}{32 \gamma ^2 \delta _b^2 p_b^4 p_c}\Big\{B^4 \gamma ^4 L_0^4 \left(\nu ^2-1\right)^2 \delta _b^4+8 B^2 \gamma ^2 L_0^2 \left(\nu ^2-1\right) \delta _b^2 p_b^2 \left(2 \xi  \cos \left(c \delta _c\right)+\xi ^2-1\right)\\
&+16 p_b^4 \left(-\gamma ^4 \delta _b^4-4 \gamma ^2 \left(\xi ^2-1\right) \delta _b^2+\left(\xi ^2-1\right) \left(4 \xi  \cos \left(c \delta _c\right)+2 \cos \left(2 c \delta _c\right)+\xi ^2-7\right)\right)\Big\}
\end{aligned}
\end{equation}
In principle, by using the Hamiltonian constraint and the fact that $p_c\sin(\delta_c c)=m\gamma L_0\delta_c$, the right hand side of the above equation can be reduced to a form with only $p_b$ and $p_c$ as arguments. This function will be denoted as $R(p_b,p_c)$ below.
A numerical result of $R$ with $\nu=0.5$ and $B=10^5$ is shown in Fig. \ref{fig:riemanncurvature}, where the classical results of $R$ are also presented as a comparison. As shown in the figure, the curvature $R$ is bounded throughout its evolution and there are two peaks which resolve the classical divergences resulted from the two classical singularities. It is remarkable that the quantum dynamics changes the sign of the curvature and the singularities with negatively infinity curvature are finally resolved by positive-valued curvature bounces.
Since $R(p_b,p_c)$ is a function with only $p_b$ and $p_c$ as arguments, one peak therefore occurs exactly at the moment $T=\tbbp$ when $p_b$ and $p_c$ bounce simultaneously.  At this moment the curvature reads
\begin{equation}
\begin{aligned}
R(\tbbp)
=\frac{2 \nu  (\nu +2)+3}{ (\nu +1)^2}\frac{8 \pi}{\gamma ^2 \Delta  }+o(\delta_b^2)\leq \frac{24 \pi}{\gamma ^2 \Delta }.
\end{aligned}
\end{equation}
which is uniformly bounded.

\section{Discussion and outlook}\label{sec:con}
In the current paper, we investigate the classical and the effective quantum dynamics of JNW spacetime. The classical dynamics indicates to us where is the classical region of the JNW spacetime, so that the quantum dynamics can be initialized. The resulted quantum dynamics matches well with the classical one in the classical region. The quantum correction leads to a bouncing evolution for both of $p_b$ and $p_c$, which means that the classical singularities can be finally resolved by the quantum theory. These results do not sensitively depend on the value of the quantum parameters.

To understand the quantum dynamics more precisely, the quantum parameters $\delta_b$ and $\delta_c$ should be fixed. In order to constraint these two parameters, we apply the idea to solve $\delta_b$ and $\delta_c$ as Dirac observables. This idea has been implemented for the Schwarzschild case, especially by the work \cite{AOS1}. However, the key equations  introduced in this work to solve the quantum parameters cannot be transported directly to our case. Thus we propose two new equations \eqref{eq:mainequations} by balancing the physical interpretation of the parameters and other physical considerations. Taking advantage of the asymptotic formula of the elliptic integrals which appears in the solutions to the Hamilton's equation, we finally obtained the approximated formula of $\delta_b$ and $\delta_c$. Properties of the results are then discussed. First of all, the values of $\delta_b$ and $\delta_c$, for $\nu=1$, return to the results for the Schwarzschild case derived by the AOS approach. Secondly, the resulted value of $\delta_c$ can be very large for small values of $\nu$, this seems to be contradictory to the usual perception that the quantum parameter should be small. However, according to our discuss, the large value of $\delta_c$ will not be a problem in the current work.

Finally, in spite of the above achievements, especially the proposal to solve the quantum parameters, there are still many attractive topics needing our further investigation. Among these topics, the most interesting one is on the extension of the resulting effective quantum spacetime. Since the integral curves of the vector filed $\frac{1}{N}\partial_T$ are geodesics with affine parameter, we can easily verify that the effective quantum spacetime with the time coordinate $T$ takes values such that $\cos(\delta_b b)\in (-1,1)$ is geodesic incomplete. It is then interesting to come up with the questions that what structures will be constructed for the maximally extended quantum spacetime and what kinds of physics will be obtained from that spacetime. All of these will be leave for our further works.

\begin{acknowledgements}
This work is supported by NSFC with No.11775082 and National Science Center (Poland) Sheng1, 2018/30/Q/ST2/00811

\end{acknowledgements}

\appendix
\section{The Elliptic Integrals}\label{app:ellip}
The three kinds of elliptic integrals are defined as
\begin{equation}
\begin{aligned}
F(\phi|k)=&\int_0^\phi\frac{1}{\sqrt{1-k^2\sin^2(\theta)}}=\int_0^{\sin(\phi)}\frac{1}{\sqrt{1-t^2}\sqrt{1-k^2 t^2}}\\
E(\phi|k)=&\int_0^\phi \sqrt{1-k^2\sin^2(\theta)}=\int_0^{\sin(\phi)}\frac{\sqrt{1-k^2 t^2}}{\sqrt{1-t^2}}\\
\Pi(n,\phi|k)=&\int_0^\phi\frac{1}{(1+n\sin^2(\theta))\sqrt{1-k^2\sin(\theta)}}=\int_0^{\sin(\phi)}\frac{1}{(1+n t^2)\sqrt{1-t^2}\sqrt{1-k^2 t^2}}
\end{aligned}
\end{equation}
where $0<k<1$. These forms are usually called the Legendre forms of elliptic integrals. A modern alternative to the Legendre forms is the carlson (symmetric) form \citep[see][for instance]{carlson1979computing}, which are defined as
\begin{equation}\label{eq:carlsonform}
\begin{aligned}
R_F(x,y,z)&=\frac{1}{2}\int_0^\infty \frac{1}{\sqrt{x+t}\sqrt{y+t}\sqrt{z+t}}\dd t\\
R_D(x,y,z)&=\frac{1}{2}\int_0^\infty \frac{1}{\sqrt{x+t}\sqrt{y+t}\sqrt{z+t}^3}\dd t\\
R_J(x,y,z,p)&=\frac{3}{2}\int_0^\infty \frac{1}{(t+p)\sqrt{x+t}\sqrt{y+t}\sqrt{z+t}}\dd t
\end{aligned}
\end{equation}
where $x,y,z\in (0,\infty)$ except that one or more of $x,y,z$ might be 0 when the corresponding integral converges. When $p<0$, the Cauchy principle value of the integral for $R_J$ is taken, which is given by
\begin{equation}\label{eq:tcauchy}
(y-p)R_J(x,y,z,p)=(\gamma-y)R_J(x,y,z,\gamma)-3R_F(x,y,z)+3R_C(xz/y,p\gamma/y)
\end{equation}
with
\begin{equation}
\gamma=y+\frac{(z-y)(y-x)}{y-p}
\end{equation}
It should be noticed that $x,y,z$ are labelled such that $0\leq x\leq y\leq z$. In addition, another useful integral $R_C(x,y)$ is defines as
\begin{equation}
\begin{aligned}
R_C(x,y):=R_F(x,y,y)=\int_0^\infty\frac{1}{\sqrt{t+x}(t+y)}\dd t.
\end{aligned}
\end{equation}
Relations between the Legendre forms and the Carlson forms are
\begin{equation}\label{eq:relationce}
\begin{aligned}
F(\phi|k)&=\sin(\phi)R_F(\cos^2(\phi),1-k^2\sin^2(\phi),1)\\
E(\phi|k)&=\sin(\phi)R_F(\cos^2(\phi),1-k^2\sin^2(\phi),1)-\frac{1}{3}k^2\sin^3(\phi)R_D(\cos^2(\phi),1-k^2\sin^2(\phi),1)\\
\Pi(n,\phi|k)&=\sin(\phi)R_F(\cos^2(\phi),1-k^2\sin^2(\phi),1)-\frac{n}{3}\sin^3(\phi)R_J(\cos^2(\phi),1-k^2\sin^2(\phi),1,1+n\sin^2(\phi))
\end{aligned}
\end{equation}
The Carlson forms possess the following properties.
\begin{itemize}
\item[(1)] Homogeneous:
\begin{equation}
\begin{aligned}
R_F(cx,cy,cz)&=c^{-1/2}R_F(x,y,z)\\
R_D(cx,cy,cz)&=c^{-3/2}R_D(x,y,z)\\
R_J(cx,cy,cz,cp)&=c^{-3/2}R_J(x,y,z,p)\\
\end{aligned}
\end{equation}
with $c,x,y,z\in \mathbb{R}^+$.
\item[(2)] Duplication theorem where we require $x,y,z,p\in\mathbb{R}^+$.
\begin{equation}
R_F(x,y,z)=2R_F(x+\lambda,y+\lambda,z+\lambda)=
R_F\left(\frac{x+\lambda}{4},\frac{y+\lambda}{4},\frac{z+\lambda}{4}\right),
\end{equation}
where $\lambda=\sqrt{x}\sqrt{y}+\sqrt{y}\sqrt{z}+\sqrt{z}\sqrt{x}$.
\begin{equation}
\begin{aligned}
R_{J}(x,y,z,p) & = 2 R_{J}(x + \lambda,y + \lambda,z + \lambda,p + \lambda) + 6 R_{C}(d^{2},d^{2} + (p - x) (p - y) (p - z)) \\
 & = \frac{1}{4} R_{J}\left( \frac{x + \lambda}{4},\frac{y + \lambda}{4},\frac{z + \lambda}{4},\frac{p + \lambda}{4}\right) + 6 R_{C}(d^{2},d^{2} + (p - x) (p - y) (p - z))
 \end{aligned}
\end{equation}
where $d = (\sqrt{p} + \sqrt{x}) (\sqrt{p} + \sqrt{y}) (\sqrt{p} + \sqrt{z})$ and $\lambda =\sqrt{x}\sqrt{y}+\sqrt{y}\sqrt{z}+\sqrt{z}\sqrt{x}$.
\end{itemize}
The duplication theorem can be used for a fast and precise computation of the Carlson form and therefore also for the computation of the Legendre forms. One can refer to \cite{carlson1979computing,press1993NRF} for more details about the numerical computation of the Carlson forms.

Asymptotic formulas of the Carlson forms when the arguments are real and tend to infinity or zero are derived in detail in  \cite{gustafson1982asymptotic}. By using the results shown in \cite{gustafson1982asymptotic}, the asymptotic behavior of $\tbb$ shown in \eqref{eq:assTbb} can be obtained as follows.
At first, when $1-k^2\ll 1$
\begin{equation}\label{eq:assRF}
\begin{aligned}
R_F(1,1-k^2,0)\cong \frac{1}{2}\ln\left(\frac{16}{1-k^2}\right)\\
\end{aligned}
\end{equation}
For $R_J(1,1-k^2,0,1+n)$, use its homogeneous to obtain
\begin{equation}
R_J(1,1-k^2,0,1+n)=\frac{1}{\sqrt{1+n}^3}R_J(\frac{1}{1+n},\frac{1-k^2}{1+n},0,1)
\end{equation}
Because the assumption that $\delta_b\ll 1$, we have
\begin{equation}
\frac{1}{1+n}\gg 1,\quad \frac{1-k^2}{1+n}=1-\nu^2+O(\delta_b^2).
\end{equation}
Then using the formula given in \cite{gustafson1982asymptotic}, which is
\begin{equation}
R_M(x,y,p)=R_M(y,x,p):=\frac{4}{3\pi}R_J(x,y,0,p)\cong \frac{4}{\pi}\frac{1}{\sqrt{y p}}R_C(p,x),\ y\to \infty
\end{equation}
we have with a straightforward calculation
\begin{equation}\label{eq:assRJ}
\begin{aligned}
R_J(1,1-k^2,0,1+n)
\cong\frac{3}{1+n}{\sqrt{\frac{n+1}{k^2+n}}}\ln\left(\frac{\sqrt{(k^2+n)}+\sqrt{n+1}}{\sqrt{1-k^2}}\right)
\end{aligned}
\end{equation}
which together with \eqref{eq:assRF} gives the asymptotic formula of $\tbb$.

\section{Discussion on solutions to \eqref{eq:Aeq}}\label{app:nogo}
As we know, $\mathcal{T}_c$ is characterized by the equation $\dd p_c/\dd T=0$, by which we have
\begin{equation}
p_c\Big|_{\mathcal T_c}=\frac{1}{2}B\nu L_0\gamma \delta_c.
\end{equation}
Then substituting the expression of $p_c\Big|_{\mathcal T_c}$ into the second equation of \eqref{eq:Aeq} gives
\begin{equation}
\delta_c\Big|_{\mathcal T_c}=\frac{\Delta}{2\pi B\nu L_0\gamma\delta_b^2\Big|_{\mathcal T_c}}
\end{equation}
which leads to the following form of the left hand side of the first equation of \eqref{eq:Aeq}
\begin{equation}\label{eq:LHS}
2\pi \delta_b\delta_c|p_b|\Big|_{\mathcal T_c}=\frac{\Delta}{2}\, \frac{ \nu\sqrt{1 -  \xi_c ^2}+\sqrt{(1-\nu^2) \gamma ^2 \delta_b^2+  \left(1-\xi _c^2\right)}}{ \nu\left(1-\xi_c^2+\gamma ^2 \delta_b^2\right)}
\end{equation}
where $\xi_c:=\xi\Big|_{\mathcal T_c}$. As there do not exist any analytical expressions for $\xi_c$, the equation \eqref{eq:LHS} can only be calculated by solving \eqref{eq:xidot} numerically with
\begin{equation}\label{eq:TofTc}
T\Big|_{\mathcal T_c}=\frac{1}{2} \ln \left(\frac{  \nu  \left(\frac{1-\nu}{1+\nu }\right)^{\nu }}{1- \nu ^2}\frac{\gamma L_0 \delta_c\Big|_{\mathcal T_c}}{B}\right)=\frac{1}{2} \ln \left(\frac{  \Delta}{2\pi B^2\delta_b^2}\frac{1}{1-\nu^2}\left(\frac{1-\nu}{1+\nu }\right)^{\nu }\right).
\end{equation}

\begin{figure}
\centering
\includegraphics[width=0.5\textwidth]{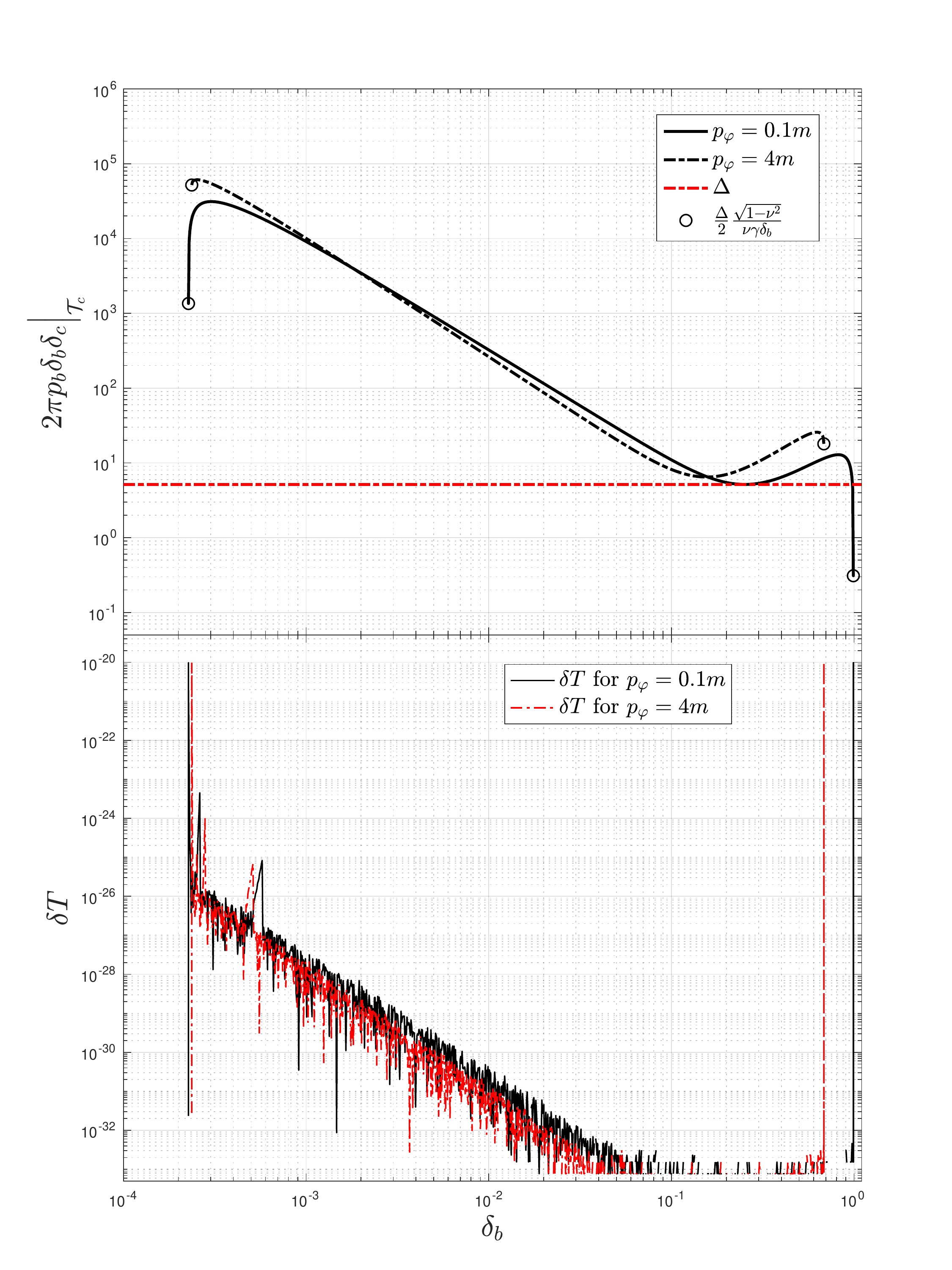}
\caption{Plot of the equation \eqref{eq:keyeq} and $\delta T$ defined in \eqref{eq:deltaT} which measures the accuracy of our computation. As shown by the plots, there do not exist any solutions to \eqref{eq:Aeq} when $p_\varphi$ is comparable with $m$, that is $p_\varphi=4m$ in this example. The parameter is chosen as $m=10^3$, $G=1=\hbar$, $\gamma=0.2375$, $L_0=1$ }\label{fig:nogoing}
\end{figure}

 Now the question becomes to solve $\delta_b$ from the equation
 \begin{equation}\label{eq:keyeq}
2\pi \delta_b\delta_c|p_b|\Big|_{\mathcal T_c}= \frac{\Delta}{2}\, \frac{ \nu\sqrt{1 -  \xi_c ^2}+\sqrt{(1-\nu^2) \gamma ^2 \delta_b^2+  \left(1-\xi _c^2\right)}}{ \nu\left(1-\xi_c^2+\gamma ^2 \delta_b^2\right)}=\Delta
 \end{equation}
 where $\xi_c$ can be computed as discussed above once $\delta_b$ is given, and therefore is a function of $\delta_b$. An example of the left- and right-hand sides of this equation  as a function of $\delta_b$ is plotted in figure \ref{fig:nogoing}, which shows to us that there does not exist any solutions when $p_\varphi=4m$ for the case $m=1000$.

There are two subtle issues here which are worth being discussed. These two issues are motivated by the following puzzle. As discussed by \cite{AOS1,AOS2}, there are four solutions to \eqref{eq:keyeq} for $\nu=1$, while the leftmost and the rightmost one are unphysical. However, when $p_\varphi$ is comparable with $m$, Fig. \ref{fig:nogoing} tells us that there do not exist any solutions to the \eqref{eq:keyeq}. Then a question arises that if the numerical results shown in Fig. \ref{fig:nogoing} are reliable  or not, which will be discussed from the following two points.

 The first of is on the range of $\delta_b$. Regarding $\xi_c$ as a function of $\delta_b$, we can obtained that
\begin{equation}
\frac{\dd \xi_c}{\dd \delta_b}=\left.\frac{\dd \xi}{\dd T}\right|_{\mathcal T_c}\frac{\dd T\big|_{\mathcal T_c}}{\dd \delta_b}<0
\end{equation}
which means that $\xi_c$ is a monotonically decreasing function on $\delta_b$. As a consequence,  $\delta_b$ ranges in $\delta_b\in [\delta_b^{+},\delta_b^{-}]$ with
\begin{equation}\label{eq:featuredeltabpm}
\xi_c\Big|_{\delta_b=\delta_b^\pm}=\pm 1.
\end{equation}
$\delta_b^+$ can be easily calculated by
\begin{equation}
T\Big|_{\mathcal T_c}=0,
\end{equation}
namely
\begin{equation}
\delta_b^+=\left(\frac{  \Delta}{2\pi B^2}\frac{1}{1-\nu^2}\left(\frac{1-\nu}{1+\nu }\right)^{\nu }\right)^{1/2}.
\end{equation}
While for $\delta_b^-$, it can only be computed numerically through the equation
\begin{equation}
T\Big|_{\mathcal T_c}=\frac{1}{2} \ln \left(\frac{  \Delta}{2\pi B^2\delta_b^2}\frac{1}{1-\nu^2}\left(\frac{1-\nu}{1+\nu }\right)^{\nu }\right)=2T^-
\end{equation}
with $T^-$ defined in \eqref{eq:initialT}. Moreover, according to \eqref{eq:featuredeltabpm}, we have that when $\delta_b=\delta_b^\pm$, the left-hand side of the key equation \eqref{eq:keyeq} becomes
\begin{equation}
\left(2\pi \delta_b\delta_c|p_b|\Big|_{\mathcal T_c}\right)\Big|_{\delta_b\to\delta_b^\pm}=\frac{\Delta}{2}\frac{\sqrt{1-\nu^2}}{\nu \gamma\delta_b}
\end{equation}
This matches with our numerical results shown in Fig. \ref{fig:nogoing}, which means that the leftmost and the rightmost solutions, existing when $\nu=1$, do disappear as shown in the figure.

The second issue is on the accuracy of our numerical computation. As discussed, $\xi_c$ can be obtained by solving the differential equation \eqref{eq:xidot} with considering the value $T\big|_{\mathcal T_c}$ given in \eqref{eq:TofTc}. This is implemented by using the 4th-order Runge Kutta method in the current paper. On the other hand, once a value of $\xi_c$ is obtained numerically, we can also calculate $T^-(\xi_c)$ by using \eqref{eq:Tofxisol}. In general, $T\big|_{\mathcal T_c}$ and $T^-(\xi_c)$ are different because of numerical errors. The difference
\begin{equation}\label{eq:deltaT}
\delta T:=\left|\left(T\big|_{\mathcal T_c}\right)-T^-(\xi_c)\right|
\end{equation}
can be used to measure the accuracy of our computation, which is shown also in Fig. \ref{fig:nogoing}. According to the results, the accuracy of our computation is $\delta T\leq 10^{-20}$. As a consequence, our computation shown in Fig. \ref{fig:nogoing} is reliable.

\end{document}